\documentclass[12pt]{article}
\usepackage{graphicx,fancyhdr,epsfig,cite}

\usepackage[pagebackref=false]{hyperref}
\usepackage{floatflt}

\begin{document}
\pagestyle{fancy}

\section*{The age specific incidence anomaly suggests that cancers originate during development.}
\begin{centering}
James P. Brody\\
Department of Biomedical Engineering\\
University of California, Irvine\\
\end{centering}

\subsection*{Abstract}

\begin{quotation}
  Cancers are caused by the accumulation of genetic alterations.
  Since this accumulation takes time, the incidence of most cancers is
  thought to increase exponentially with age.  However, careful
  measurements of the age-specific incidence shows that the specific
  incidence for many forms of cancer rises with age to a maximum, then
  decreases.  This decrease in the age-specific incidence with age is
  an anomaly. Understanding this anomaly should lead to a better
  understanding of how tumors develop and grow.  Here I derive the
  shape of the age-specific incidence, showing that it should follow
  the shape of a Weibull distribution.  Measurements indicate that the
  age-specific incidence for colon cancer does indeed follow a Weibull
  distribution.  This analysis leads to the interpretation that for
  colon cancer two sub-populations exist in the general population: a
  susceptible population and an immune population.  Colon tumors will
  only occur in the susceptible population.  This analysis is
  consistent with the developmental origins of disease hypothesis and
  generalizable to many other common forms of cancer.
\end{quotation}

\subsection*{Introduction}

\begin{figure*}[tbh] \centering 
\includegraphics[width=5in]{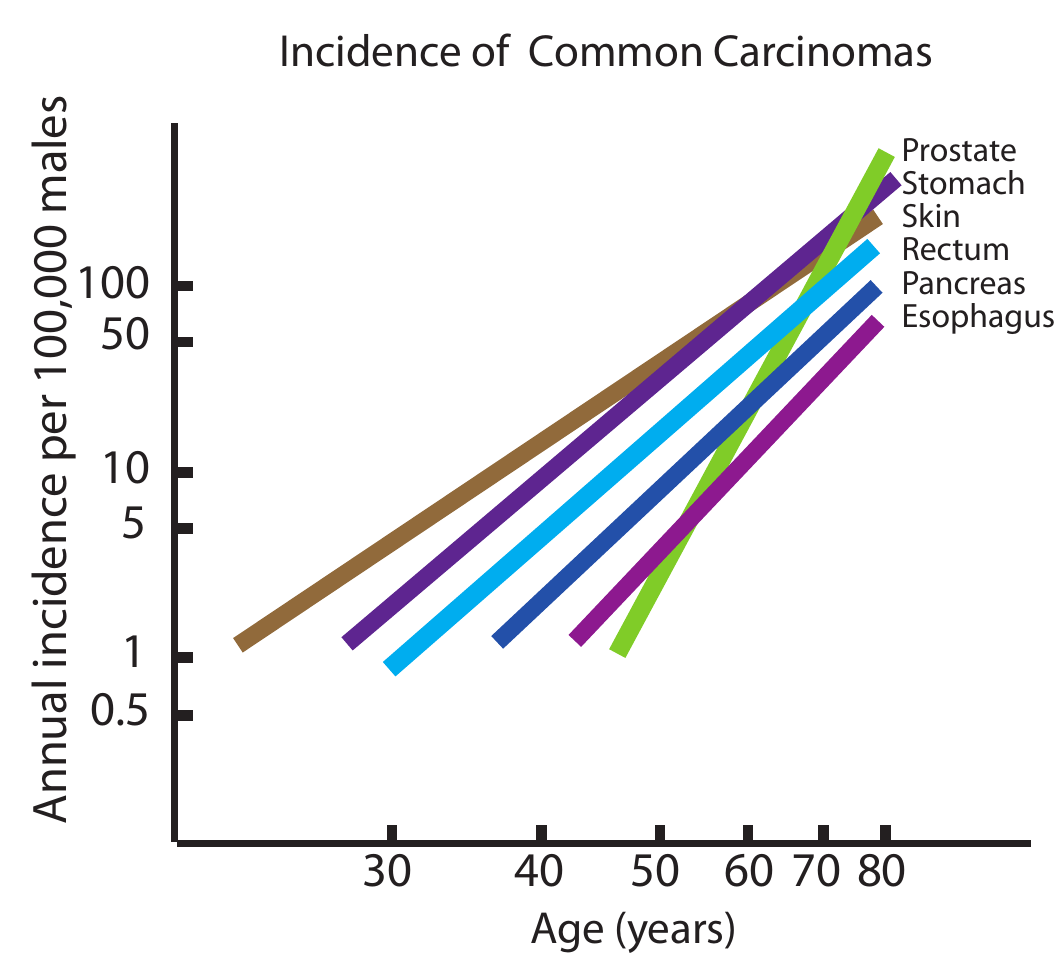}
\caption[]{The textbook illustration of how cancer incidence increases
  with age.  The specific incidence of cancers is often depicted as
  exponentially increasing with age.  An illustration similar to this
  appears in a popular undergraduate molecular biology
  textbook\cite{lodish2004}.  } \label{oldincidence}
\end{figure*}

Cancers are thought to originate after a series of genetic alterations
accumulate in a cell.  These alterations could consist of mutations,
deletions or modifications to the DNA. The accumulation of alterations
increases when certain pathological states occur: chromosomal
instability, DNA repair defects. A typical colorectal cancer genome
contains about a dozen mutated genes that are considered to be driving
the cancer\cite{Sjoeblom2006,Wood2007}. Since a normal cell needs to
accumulate mutations to more than a dozen key genes before
transforming into a tumor cell\cite{Markowitz2009}, the probability of
acquiring a particular cancer should increase with age.  Thus, it is
widely thought that the older one gets, the more likely one is to
develop cancer. Age is the primary risk factor for cancer, and cancer
is considered an age related disease \cite{Campisi2009}.  The textbook
understanding of how age relates to cancer is shown in
\autoref{oldincidence}.

However, the incidence of most cancers does not monotonically increase
with age; instead, the incidence increases to a maximum at some age
and then decreases. Although this anomaly
is well established \cite{Harding2012,pompei2001a,Harding2008}, it is not
widely known.

A complete understanding of the age-specific incidence should lead to
a better understanding of how cancers develop.  The age-specific
incidence is the only quantitative data available on the development
process of cancer.  It is not confounded by animal models.  One of the
primary steps to understanding the age-specific incidence is to
understand the anomaly.

\subsection*{Thought experiments}
We can begin with a \emph{gedanken} experiment.  Take 100,000 newborn
human infants and put them in an isolated box.  These babies live well, and
grow into adults.  In this idealized experiment, they do not die of any ailments.  Each of these experimental
subjects is regularly examined for a particular type of cancer, say
colon cancer.  When a subject is first diagnosed with a colon
cancer, the exact age is recorded.  The experiment runs for hundreds
of years, then we make a histogram of the number of 
colon cancers diagnosed as a function of years since the beginning of
the experiment.

\begin{figure*}[tbh] \centering 
\includegraphics[height=4in]{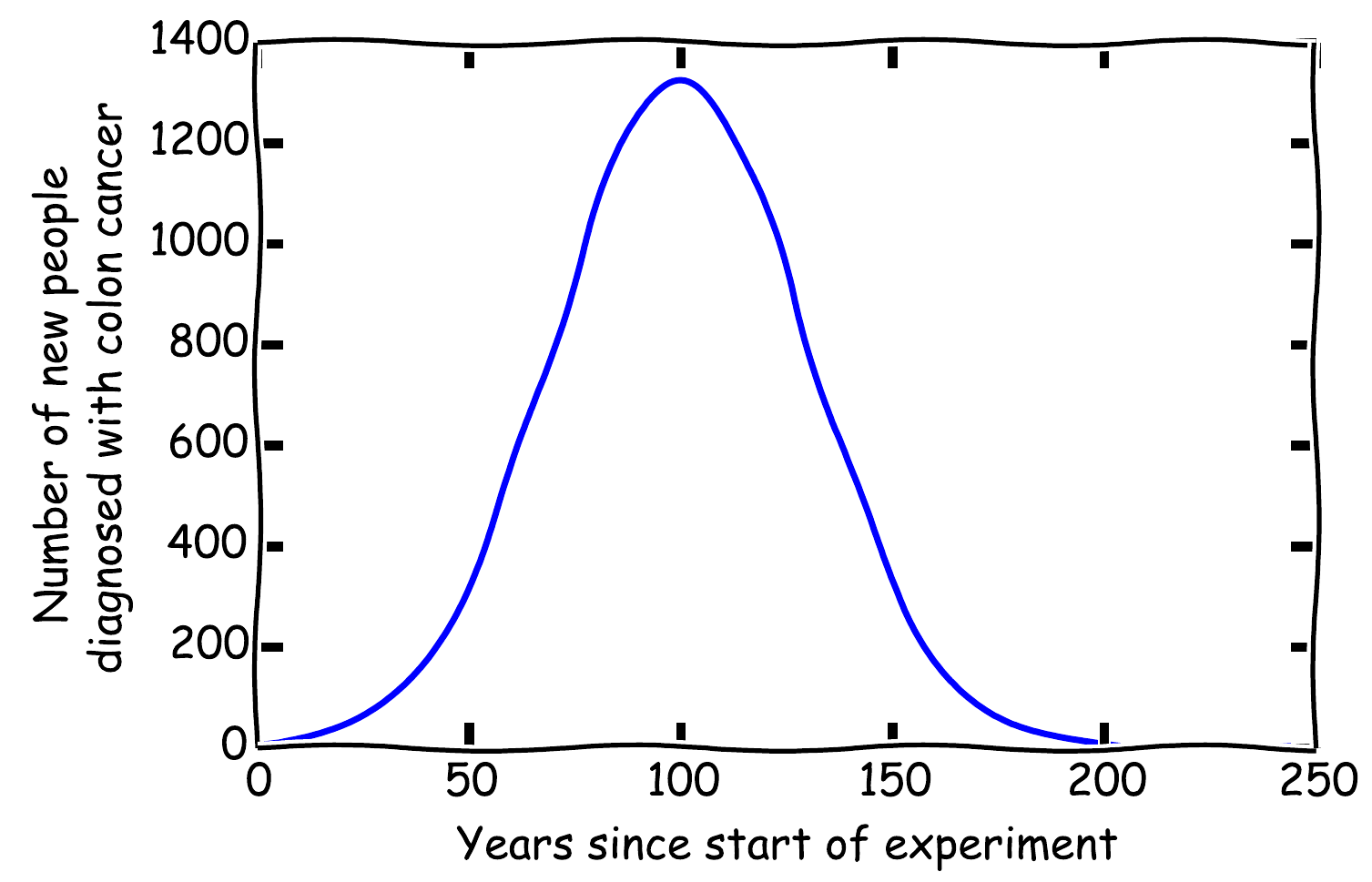}
\caption[]{The postulated results of a simple thought experiment. The
  experiment consists of 100,000 newborn humans isolated and monitored
  annually for colon cancer.  The number diagnosed with colon cancer
  for the first time is recorded each year. After several hundred
  years, the data is plotted.  The plot shows that the incidence
  increases, reaches a maximum, then decreases to zero. The decrease
  occurs when the majority of the population has already been
  diagnosed once with colon cancer. } \label{gedanken}
\end{figure*}

This histogram will start near zero, reach a maximum, then decline to
zero again when all subjects have been diagnosed with a first case of
colon cancer, as shown in \autoref{gedanken}.  If every member of the
initial population of infants ultimately developed the cancer, then
the integral of this histogram should be equal to the population, in
this case, 100,000.

A second \emph{gedanken} experiment is also performed.  In this case,
the population of 100,000 infants is composed of two apparently
indistinguishable sub populations, one of which can develop colon
cancer (20\%), and the other of which cannot(80\%).  The same process
is followed.  At the end of the experiment, the integral of the
histogram is calculated and it will equal 20,000.

While these \emph{gedanken} experiments are unrealistic, an analogous
experiment can be done. First, we record the age of all patients
diagnosed with a specific cancer within a large geographic area in one
year.  Second, we record the age of each member in the entire
population.  Finally, for each age group, we divide the number of
patients who had a tumor diagnosed in that year by the total number of
people with that age in the population.  By convention, these numbers
are multiplied by 100,000 and are called the age-specific incidence.

\subsection*{Population based cancer registries}
Population based cancer registries record the age (and other
information) about all patients diagnosed with all types of tumors
within a specific geographic area.  Then, a government census records
the ages of all members of the population within a specific geographic
area.  Together these sources of data can be combined to compute the
age-specific incidence data.

The collection and quality of age-specific incidence data has
significantly improved since registries began in the mid 1900's.  Initially, this data was
derived from death certificates.  However, many deaths were attributed
to ``old age'' or nonstandard terminology. Today, cancer registries systematically collect
information on the diagnosis of tumors and demographic information of
the patient.

Different cancer registries, however, collect different information.
These differences make aggregation of cancer registry data difficult. In 1973,
the National Cancer Institute established the Surveillance, Epidemiology, and End Results (SEER) program.  The SEER network of cancer registries  solved many of these problems by requiring a
specific set of information to be reported and established guidelines
on how to encode different properties of a tumor.  

The SEER network of cancer registries began in 1973 with seven
different geographic registries covering 16 million people.  The program has
expanded from seven cancer registries in 1973 to 18 cancer registries in
2012, with about 86 million people under surveillance.  The SEER
program publishes annually case files, which contain summary
information about all tumors diagnosed within the specific geographic
areas.

\begin{figure*}[tbh] \centering 
\includegraphics[height=6in]{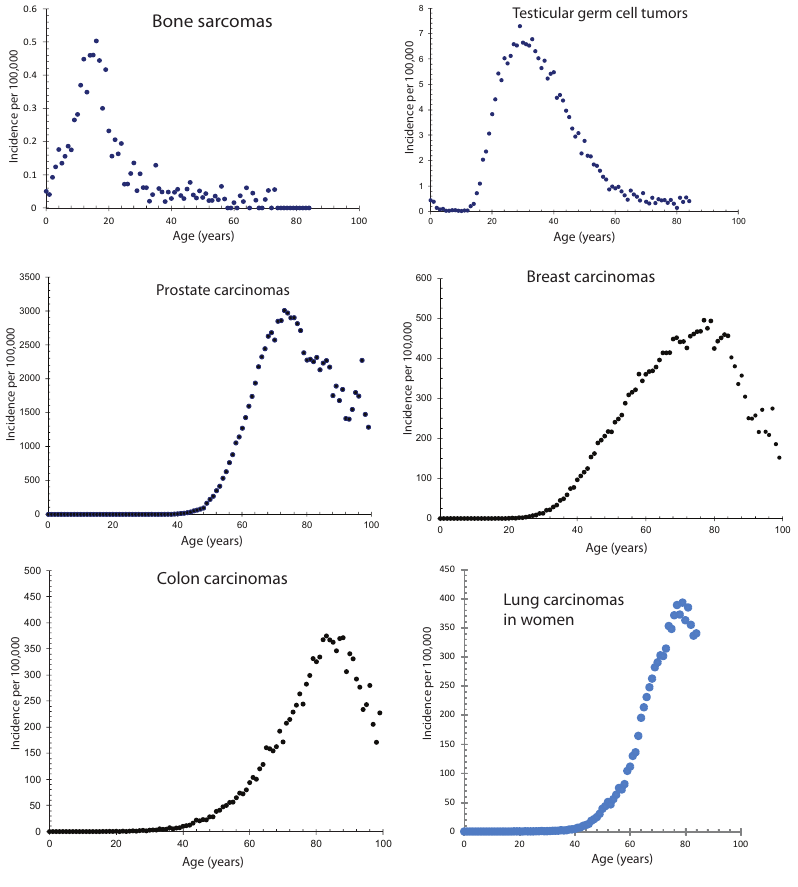}
\caption[]{Many different forms of cancer exhibit the age-specific
  incidence anomaly.  This figure shows that cancers of the bone,
  testicles, prostate, breast, colon, and lung all increase with age,
  reach a maximum, then decrease.  Understanding this anomalous
  decrease in the age-specific incidence should lead to a better
  understanding of how cancers
  begin.} \label{manycancers} \end{figure*}

Age-specific incidence data collected by the SEER-17 network of cancer
registries in 2000 is shown in \autoref{manycancers}.  This data is presented to
emphasize that different cancers have different maximum ages.  In each
of the seven cancers shown, a decrease in incidence with age exists.
This decrease is anomalous, the opposite of the expected behavior, but
is consistent with our \emph{gedanken} experiments.

\subsection*{Anomaly or Artifact?}

A number of concerns have been raised with the surprising observation
that the age-specific incidence decreases with age. The three most
common are (1) this observation is contradicted by autopsy studies of
latent carcinoma, (2) this observation is an artifact caused by
decreased screening rates with age, and (3) this observation is the
result of a birth cohort effect.  Each of these concerns has been
studied in great detail and none of these are sufficient to explain
the anomaly.

\textbf{Autopsy studies of latent carcinoma.} A widespread perception
exists that undiagnosed carcinomas, or latent carcinomas,  are common in elderly people.  Much
of this perception is due to work done in the 1950's by LM Franks, who
performed several autopsy studies \cite{FRANKS1954,FRANKS1954a}.  His work
appeared to show that many undiagnosed cancers existed in people who
died.  This work was based on rather small numbers (for instance,
\cite{FRANKS1954a} only had two subjects in the 90-100 year old age
group.)  This data has not been replicated, moreover recent autopsy
studies on larger populations \cite{Imaida1997,derijke2000} contradict
Frank's results and have established that the incidence of many common
carcinomas \emph{decreases} after a certain age.

\textbf{Screening and age-specific incidence data.} The effect of
cancer screening on colorectal cancer rates can be estimated.
Colorectal screening rates decrease with age after 60 years.  The
National Survey of Ambulatory Surgery quantified the rate of
outpatient colonoscopies (over 90\% of colonoscopies are performed as
outpatients) in 1994, 1995, 1996 and 2006\cite{Cullen2009}.  (The
survey was not performed in the years between 1996 and 2006.) Based on
these estimates, colorectal screening rates in the elderly population
(over 85) were  about 40\% of the rate of 50 to 64 year olds.

The increase in diagnosed cancers due to screening can be estimated
from the age-specific incidence data.  Guidelines suggest beginning
screening at 50 years of age.  The colon carcinoma age-specific
incidence data shows a small, but noticeable, increase over the
expected rate at 50 years of age.  From this, we estimate the number
of new cases of colon carcinoma due to screening at about 2 per
100,000, when about 5000 per 100,000 are screened.  Based on these
numbers, we estimate that if screening rates did not decrease with
age, the specific incidence of colon carcinoma would increase by about
40 cases per 100,000 population at age 85.  This is not a significant
difference.

\textbf{Birth cohort effects on age-specific incidence data.} The drop
in the incidence of colon cancer after age 85 is not due to birth
cohort effects.  The expected value, if 100\% of the population were
susceptible, for age 99 is about 850 per 100,000, as shown in data
from 200 shown \autoref{weibull-theory}.  The observed value is about
227 with a 95\% confidence interval of 96 to 357.  The observed value
is about one quarter of the expected value of colon carcinoma
incidence, if 100\% were susceptible to colorectal carcinoma.  If this
population, born in 1900-1910, had a significantly reduced propensity
to develop colorectal carcinoma, then we should see a correspondingly
small incidence in the population aged 72 years old recorded in 1973
(the earliest SEER data available). No such effect is noticeable in
the 1973 data.

Finally, one other objection to the observation of declining cancer
incidence with age is sometimes raised mistakenly: competing
risk. Competing risk is not relevant here. It is relevant to studies
with fixed populations when some members of the population die from
other causes. The age-specific incidence data is not based on fixed
populations.  Specific incidence is the number of diagnosed cancers
divided by the population.

\subsection*{Understanding the  age-specific incidence data}

We follow two approaches to understand the age-specific incidence
data.  The first approach is theoretical: based on first principles,
what should be the shape of the age-specific incidence curve?  The
second approach asks what biomedical hypothesis could produce
age-specific incidence data that we observe.

\textbf{A theory of the age-specific incidence curve}.  We have
postulated that the age-specific incidence curve should follow a
extreme value distribution, in particular the Weibull
distribution\cite{Sotoortiz2012}.  Our reasoning is that:
\begin{enumerate}
 \item Tumors originate in a single cell, the progenitor cell. 
 \item  Many potential progenitor tumors cells exist in the body for each type of potential tumor.  
   \item A tumor develops when the \textbf{first} of  these many potential progenitor cells.
\end{enumerate}
These steps describe an extreme value process.  The Weibull
distribution is the proper distribution to characterize this
process\cite{WEIBULL1951}.

The probability of developing a particular cancer as a function of
time, $p(t)$, is given by the Weibull distribution
\begin{equation}
p(t)=A \left(\frac{k}{\lambda}\right) 
       \left(\frac{t-\tau}{\lambda}\right)^{k-1}
       \exp^{-(\frac{t-\tau}{\lambda})^k},
\label{weibull}
\end{equation}
when $t\ge \tau$ and the Weibull distribution is $p(t)=0$ when $t<
\tau$.  The Weibull distribution has four parameters: $A$ is a
normalization factor, $\tau$ is the time shift, $k$ is called the
shape parameter, and $\lambda$ is the scale parameter.  Both $k$ and
$\lambda$ must be positive.

\begin{figure*}[tbh] \centering 
\includegraphics[width=5in]{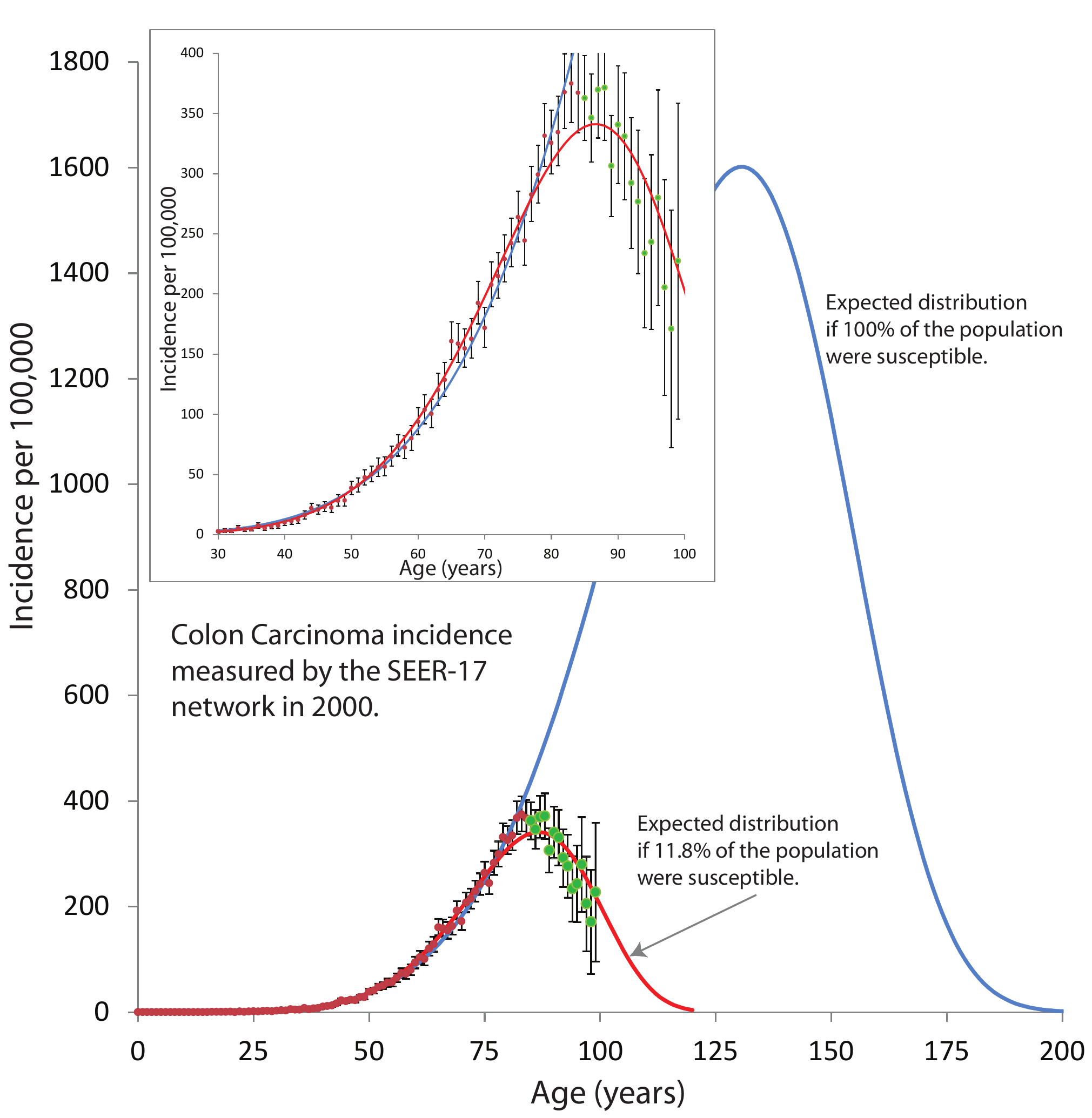}
\caption[]{The age-specific incidence of colon carcinoma, as measured
  by the SEER-17 network of cancer registries in 2000, is plotted
  along with the best fit Weibull distribution for 100\% of the
  population and for 11.8\% of the population.  Both distributions
  were fit to data from ages 0 to 85 years of age, then data from
  86-99 years of age were plotted. The data is clearly consistent with
  the 11.8\% curve, but not the 100\% curve.  } \label{weibull-theory}
\end{figure*}

We compared the theoretical shape of the age-specific incidence data
with observational data collected by the SEER-17 registries in 2000
for colon cancer.  We computed the best fit theoretical shape to the 0-84
year old data  in two cases.  First, when all three parameters
($A,k,\lambda$) were allowed to vary and second when $A$ was fixed to
be 100,000, but the other two parameters were allowed to vary.  These
results are shown in \autoref{weibull-theory}.

The two fits in \autoref{weibull-theory} correspond to the two
\emph{gedanken} experiments.  If everyone eventually will develop
colon cancer, the data points should fall on the 100\% line.  We
observe the 86-100 years old data points, which were not involved in
the fitting, to fall on the 11.8\% curve.

This analysis suggests that two subpopulations exist.  One
subpopulation, consisting of about 12\% of the population, is
susceptible to developing colon cancer.  The second subpopulation,
consisting of about 78\% of the population is immune to developing
colon cancer.  Membership in the susceptible population must be
determined early, before the age of 20.

One objection to this interpretation is that it apparently contradicts
the well-established observation that modifiable risk factors exist
for most common cancers.  The link between environmental exposure and
increased cancer rates is well established, most prominently between
cigarette smoking and lung cancer.  If environmental exposure causes
cancer, how can a susceptible subpopulation exist and be defined at an
early age?  

One possible explanation is that environmental risk factors effect
how fast a tumor grows.  For instance, a non-smoker predisposed to
lung cancer might develop a lung tumor at age 120, while a similar
heavy smoker develops a tumor at age 60.  Since the non-smoker will
probably die from other causes before the lung tumor develops, it
appears that the smoker developed lung cancer while the non-smoker did
not develop lung cancer. 

The idea that disease observed late in life could originate early in
life is not novel. This idea is called the developmental origin of
disease hypothesis\cite{Barker1990,Barker2003b,Barker2002}

\textbf{Developmental origin of disease hypothesis.} Different forms
of this hypothesis have been
proposed\cite{Morgenthaler2004,Herrero-Jimenez1998}.  Trichopoulos has
suggested that hormonally regulated cancers originate \emph{in utero}.
He points out that this would explain a number of curious observations
about breast cancer including the dramatic difference in incidence
found in Japan and the USA \cite{Trichopoulos1990}. Barker has
suggested that not only cancers, but also other adult diseases have
fetal origins \cite{Barker1990,Calkins2011}. Others have also
suggested that some chronic diseases are influenced by exposure to
environmental factors early in life
\cite{Gluckman2004b,Gluckman2008b}.  Diabetes\cite{Yajnik2004b},
schizophrenia\cite{Stclair2005}, and lung disease\cite{Harding2012a}
might also find their origins in early life.

\textbf{Mechanisms for the developmental origins of disease
  hypothesis.} Several known mechanisms could be responsible for the
existence of two sub populations, these include germ line mutations,
somatic mutations early in life, and/or epigenetic modifications
inherited or acquired early in life.

Germ line mutations have been ruled out.  During the 1990's,
significant resources were devoted to the identification of germ line
mutations for the most common forms of cancers.  This effort led to
the identification of BRCA1\cite{Miki1994}.  Certain mutations in
BRCA1 significantly increase the risk that a woman will develop breast
cancer.  However, these mutations are rare and less than 10\% of
breast cancers in the US population occur in women with these
mutations. Despite searching for similar genes in colon cancer
\cite{Peltomaeki1993}, none have been found with the significance of
BRCA1. No recurrent mutations are responsible for the progression of
colon cancer \cite{Feinberg2006}.

Somatic mutations acquired early in life (during development) could
propagate to encompass entire tissues. Embryonic cells are actively
proliferating and a somatic mutation acquired early during development
will be found in many cells. Irradiation of a fetus is
known to increase the incidence of childhood cancers\cite{Doll1997}
presumably through the acquisition of somatic mutations. Somatic
mutations acquired during development are known to be responsible for
retinoblastoma, a childhood cancer \cite{Frank2003}.

Epigenetic alterations play a key role in the carcinogenesis
process\cite{Esteller2008,Jirtle1999,Jirtle2007}.  These types of
alterations can be passed down through cellular generations.  Modification of
histones are a key regulatory step in transcription \cite{Jones2007}
and DNA damage repair \cite{Chi2010}. Specific histone modifications
have been identified that are common features of human cancers
\cite{Fraga2005,Das2009}.  Several approaches to determining genome
wide methylation exist, but these approaches have not yet been widely
applied to cancer  \cite{Laird2010}.

\subsection*{Conclusion}

In conclusion, we showed that logically (through \emph{gedanken} experiments), theoretically, and observationally that the  age-specific incidence data decreases with age.  This apparent anomaly is consistent with the developmental origin of disease hypothesis.

\subsection*{References}

\renewcommand\refname{}


\bibliographystyle{naturemag}

\end{document}